\documentclass[journal=jacsat,manuscript=article,layout=twocolumn,psfig,pstricks]{achemso}

\setkeys{acs}{super=true}
\setkeys{acs}{articletitle=true}
\setkeys{acs}{chaptertitle=true}
\setkeys{acs}{etalmode=truncate,maxauthors=0}
\setcitestyle{super,open={},close={}}

\usepackage{amsfonts}
\usepackage{amsmath}
\usepackage{amssymb}
\usepackage[english]{babel}
\usepackage{bm}
\usepackage{booktabs}
\usepackage{cancel}
\usepackage[format=plain,singlelinecheck=false,font={small,bf},labelfont=bf,labelsep=period]{caption}
\usepackage{cleveref}
\usepackage{colortbl}
\usepackage{csquotes}
\usepackage{dcolumn}
\usepackage{fancyhdr}
\usepackage{float}
\usepackage{geometry}
\usepackage{graphicx}
\usepackage{helvet}
\usepackage{hyperref}
\hypersetup{breaklinks=true,colorlinks=true,citecolor=blue,linkcolor=blue,filecolor=blue,urlcolor=blue}
\usepackage{listings}
\usepackage{lscape}
\usepackage{mathpazo}
\usepackage{mathtools}
\usepackage[version=3]{mhchem}
\usepackage{microtype}
\usepackage{multirow}
\usepackage{natbib}
\usepackage{natmove}
\usepackage{pgfplots}
\usepackage{pstricks}

\usepackage[fontsize=10.0pt]{scrextend}
\usepackage{sectsty}
\usepackage{setspace}
\usepackage{siunitx}
\usepackage{subfigure}
\usepackage{tablefootnote}
\usepackage[compact]{titlesec}
\usepackage{txfonts}
\usepackage[normalem]{ulem}
\usepackage{xcolor}
\usepackage{xkeyval}
\usepackage{color}
\usepackage[normalem]{ulem}

\def\SM{Supplementary Materials}
\def\TBP{trigonal-bipyramidal}

\SectionsOn
\SectionNumbersOn
\AbstractOn
\author{Diego Guedes-Sobrinho}
\email{guedes.sobrinho.d@gmail.com}
\affiliation[University of S\~ao Paulo]{S\~ao Carlos Institute of Chemistry, University 
of S\~ao Paulo, PO Box $780$, $13560-970$, S\~ao Carlos, S\~ao Paulo, Brazil}

\author{Weiqi Wang}                                                             
\email{wang@fhi-berlin.mpg.de}                                                  
\affiliation[Fritz-Haber-Institut der Max-Planck-Gesellschaft]{Fritz-Haber-Institut 
der Max-Planck-Gesellschaft, $14195$ Berlin-Dahlem, Germany}

\author{Ian Hamilton}
\email{ihamilton@wlu.ca}
\affiliation[Wilfrid Laurier University]{Department of Chemistry and Biochemistry, Wilfrid Laurier 
University, Waterloo, N2L 3C5 Ontario, Canada}

\author{Juarez L. F. Da Silva}                                                  
\email{juarez_dasilva@iqsc.usp.br}                                              
\affiliation[University of S\~ao Paulo]{S\~ao Carlos Institute of Chemistry, University 
of S\~ao Paulo, PO Box $780$, $13560-970$, S\~ao Carlos, S\~ao Paulo, Brazil}

\author{Luca M. Ghiringhelli}                                                   
\email{ghiringhelli@fhi-berlin.mpg.de}                                          
\affiliation[Fritz-Haber-Institut der Max-Planck-Gesellschaft]{Fritz-Haber-Institut 
der Max-Planck-Gesellschaft, $14195$ Berlin-Dahlem, Germany}

\title[(Meta-)stability and Core-Shell Dynamics of Gold Nanoclusters at Finite Temperature]{(Meta-)stability and Core-Shell Dynamics of Gold Nanoclusters at Finite Temperature}

\abbreviations{DFT, PBE, FHI-aims, BOMD}
\date{\today}

\begin{document}
\maketitle

\begin{abstract}
Gold nanoclusters have been the focus of numerous computational studies but an atomistic understanding of their structural and dynamical properties at finite temperature is far from satisfactory. To address this deficiency, we investigate gold nanoclusters via {\em ab initio} molecular dynamics, in a range of sizes where a core-shell morphology is observed. We analyze their structure and dynamics using of state-of-the-art techniques, including unsupervised machine-learning nonlinear dimensionality reduction (sketch-map) for describing the similarities and differences among the range of sampled configurations. We find that, whereas the gold nanoclusters exhibit continuous structural rearrangement, they clearly show persistent motifs: a cationic core of one to five atoms is loosely bound to a shell which typically displays a substructure resulting from the competition between locally spherical vs planar fragments. Besides illuminating the properties of core-shell gold nanoclusters, the present study proposes a set of useful tools for understanding their nature in {\em operando}.
\end{abstract}

\section*{}
Transition-metal nanoclusters have been the focus of a large number of experimental and computational studies in physics and chemistry, due to their expected technological applications in diverse areas, including catalysis,\cite{Aiken1999} optics,\cite{Jin2015} and biomedicine.\cite{Marjomaki_1277_2014}
Their physical and chemical properties are dependent on many characteristics such as size,\cite{Chaves_15484_2017} shape,\cite{Mostafa_15714_2010} charge state,\cite{Chaves_10813_2014} ligand effects,\cite{Jadzinsky_430_2007} and temperature\cite{Yamauchi_8388_2013}.
Transition-metal nanoclusters typically exhibit catalytic activity at high temperature, \cite{Sugimura1998} but gold nanoclusters have recently received special attention due to the discovery of their catalytic activity at low temperature.\cite{Haruta_75_2003}

Experimental and computational studies of both colloidal \cite{Haekkinen_3264_1999,Heaven_3754_2008,Farrag_12539_2013,Azubel_909_2014,Jin_10346_2016} 
and bare \cite{Garzon2_105_2003,Xing_165423_2006,bulusu2007structural,Jiang_193402_2011,Wang_4038_2012,Zeng_2976_2015,shao2014structural,Pande_10013_2016} gold nanoclusters have achieved important advances in the last decades. 
Most of the reported {\em ab initio} studies have been dedicated to searching for their putative global (energy) minimum configurations (pGMC) in the gas phase at zero temperature. \cite{Azubel_909_2014}
An empirical-potential molecular-dynamics (MD) study of \ce{Au75}, \ce{Au146}, and \ce{Au457}, showed that their melting temperatures were well below the melting temperature of bulk gold (\SI{1090}{\kelvin}, at standard pressure) but also that structural solid-to-solid transitions were present well below their melting temperatures.\cite{Cleveland_5065_1999} For instance, for \ce{Au75}, the melting temperature is around \SI{550}{\kelvin}, but around \SI{350}{\kelvin} there is a structural transition from truncated-decahedral to icosahedral. 

The present study focuses on the gas-phase gold nanoclusters \ce{Au25}, \ce{Au38}, and \ce{Au40}.  
\ce{Au25} was chosen because it is expected to be the smallest cluster that has a core --- in this case a single Au atom --- in the low-lying isomer population.\cite{bulusu2007structural}
\ce{Au38} and \ce{Au40} were chosen because they are both expected to have tetrahedral \ce{Au4} cores and low-symmetry structures have been proposed as their pGMCs.\cite{shao2014structural,Pande_10013_2016}
We use \textit{ab initio} Born-Oppenheimer MD, an important computational technique that allows for the sampling of the configurational space of systems with many atoms, together with yielding reliable information on the dynamics of the system, e.g., in terms of time-correlation functions. When the Born-Oppenheimer MD equations of motion are extended by adding a thermostat, the canonical ensemble (at temperature $T > \SI{0}{\kelvin}$) is sampled. In the specific case of density-functional-theory (DFT) based MD, the energy of the system and forces between atoms are determined by solving the Kohn-Sham equations. Hence, the ground-state electronic structure is accessible along the simulated trajectories.

DFT-based MD studies on small Au clusters have been performed in the past, aimed at describing the thermal evolution of the ground-state structures and the competition between structural isomers at a given temperature,\cite{Bas_55_2005,Kang_18287_2010,Beret_153_2011} as well as at accessing the infra-red (vibrational) spectra for comparison with experiments.\cite{Ghiringhelli_083003_2013}
We analyze DFT-based MD trajectories of \ce{Au25}, \ce{Au38}, and \ce{Au40} at room and higher temperatures, in order to address important questions regarding the structural and dynamical aspects specific to this range of cluster sizes:
$(i)$ is the core-shell morphology persistent at finite temperature? $(ii)$ do the atoms constituting 
the core exchange their positions with those constituting the shell? $(iii)$ 
does the shell of different isomers have recurrent structural motifs?

\begin{figure*}[t!]
\centering
\includegraphics[width=0.8\textwidth]{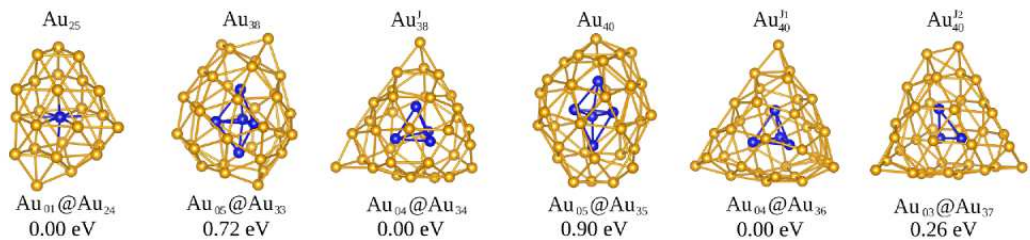}
\caption{Relative total energies and configurations used as starting points of the MD trajectories. Structures labeled as Au$_{38}^\textrm{\,J}$, Au$_{40}^\textrm{\,J1}$, and Au$_{40}^\textrm{\,J2}$
are from Ref.\citenum{Jiang_193402_2011}, while structures labeled as \ce{Au25}, \ce{Au38}, and \ce{Au40} are found via EAM-RBHMC and further optimized with PBE+MBD (see text for details). 
The core (shell) atoms are indicated by blue (yellow) colour. The labeling Au$_c$@Au$_s$ marks the number of atoms in the core ($c$) and shell ($s$).}
\label{initial_structures}
\end{figure*}

We ran DFT-based MD within the Perdew--Burke--Ernzerhof\cite{Perdew_3865_1996} (PBE) approximation for the electronic exchange-correlation functional,
corrected for the many body dispersion (MBD) interactions\cite{Tkatchenko_074106_2013}, in the all-electron full-potential implementation of the \textsc{FHI-aims} package. \cite{Blum_2175_2009}
The time step for the numerical integration of the MD equations of motion was set to \SI{10}{f\second}
and we applied a Gaussian broadening  of \SI{1}{\milli\electronvolt} for the occupation at the Fermi level.
Canonical sampling was imposed via the stochastic velocity-rescaling thermostat.\cite{Bussi_014101_2007}
Further MD details are described in the \SM.
These settings have been already adopted in Refs. \citenum{Beret_153_2011,Ghiringhelli_083003_2013,GhiringhellI_JCP_2017} and shown to yield observable quantities in good agreement with
experiments in the size range between 3 and 20 gold atoms. Gold nanoclusters are strongly influenced by relativistic effects,\cite{Haekkinen_093401_2004,Huang_987_2008} as the contraction of the $6s$ orbital and expansion of the $5d$ results in a smaller $s-d$ gap and an increased directionality in the Au--Au bonding. In the present study, relativistic effects are included via the `atomic' zeroth-order regular approximation to the Dirac Equation.\cite{van2000gradients}

The initial configurations for the MD trajectory were (Fig. \ref{initial_structures}): 
the pGMC structures described in Ref. \citenum{Jiang_193402_2011} for \ce{Au38} and \ce{Au40}, both containing a tetrahedral \ce{Au4} core,
and a higher energy isomer for \ce{Au40} (also from Ref. \citenum{Jiang_193402_2011}), containing an \ce{Au3} core.
Furthermore, to increase diversity in the starting point of the MD trajectories of \ce{Au38} and \ce{Au40}, we also selected the respective pGMC of a potential-energy surface described through embedded-atom method (EAM), globally optimized via the revised basin-hopping Monte Carlo (RBHMC) algorithm,\cite{Rondina_2282_2013} both containing a \ce{Au5} core.
For \ce{Au25}, we adopted the embedded-atom-RBHMC pGMC, which has a one-atom core. More details on the RBHMC search are given in Fig. S1 of the \SM.
For each starting point, one or more \textit{ab initio} MD trajectories of \SI{100}{\pico\second} is run, where the first interval of \SI{50}{\pico\second} is thermostatted at a temperature between 400 and \SI{600}{\kelvin} (see Table S1 of the \SM\ for details on the single trajectories) and the last interval of \SI{50}{\pico\second}  is at `room temperature', i.e., \SI{300}{\kelvin}.

\begin{figure*}[t!]
\centering
\includegraphics[width=0.75\textwidth]{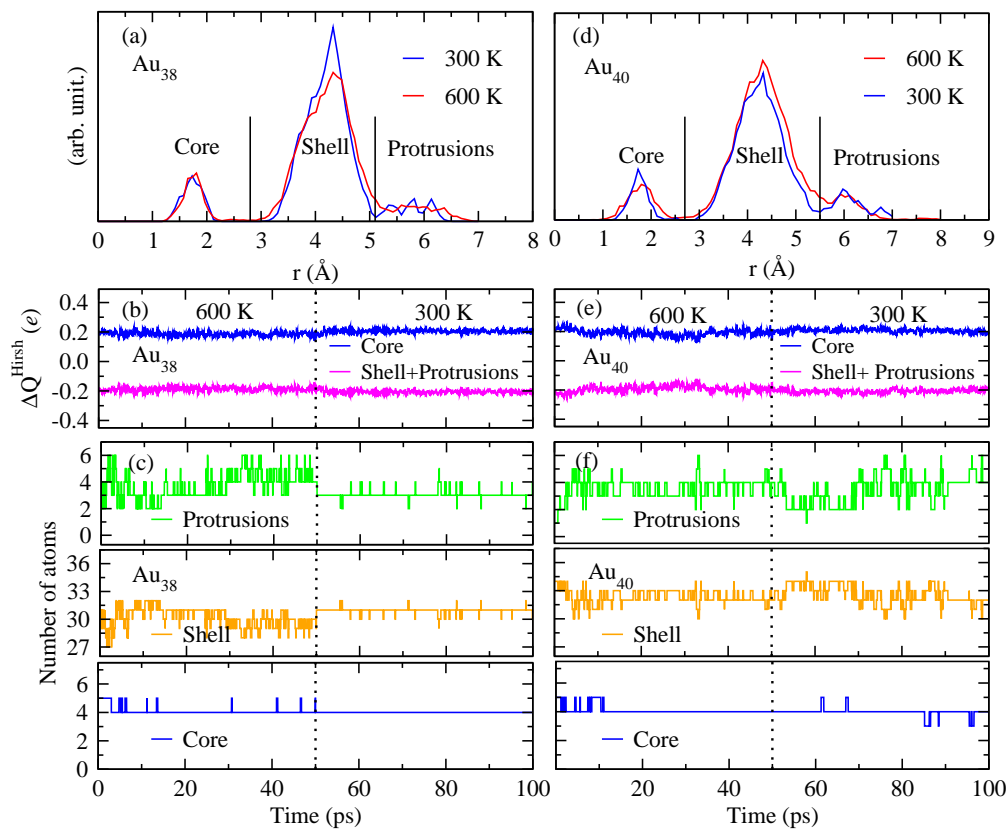}
\caption{(a) Radial distribution function, $g(r)$, and cutoff distances (solid vertical lines) with respect to the cluster's center of mass defining the core, shell, and protrusions regions for the \ce{Au38} nanocluster at \SI{600} and \SI{300}{\kelvin}; (b) evolution of the Hirshfeld charges for core and shell atoms and (c) evolution of number of atoms belonging to protrusions (green line), shell (orange line), and core (blue line) regions at \SI{600} and \SI{300}{\kelvin} for \ce{Au38}. Panels (d)--(f) respectively show the same quantities as in (a)--(c), but for \ce{Au40}.}
\label{hirshfeld_radial_number}
\end{figure*}

As detailed below, the statistical analysis of the structure and dynamics of the studied clusters is performed by means
of radial-distribution functions, bond-lifetime analysis, nonlinear dimensionality reduction (in order to represent, by exploiting ``structure similarities'', the high-dimensional configurational space in two dimensions), and state-of-the-art Boltzmann re-weighting (in order to use the MD sampling at all temperatures in the most efficient way).

\begin{figure*}[t!]
\centering
\includegraphics[width=0.95\textwidth]{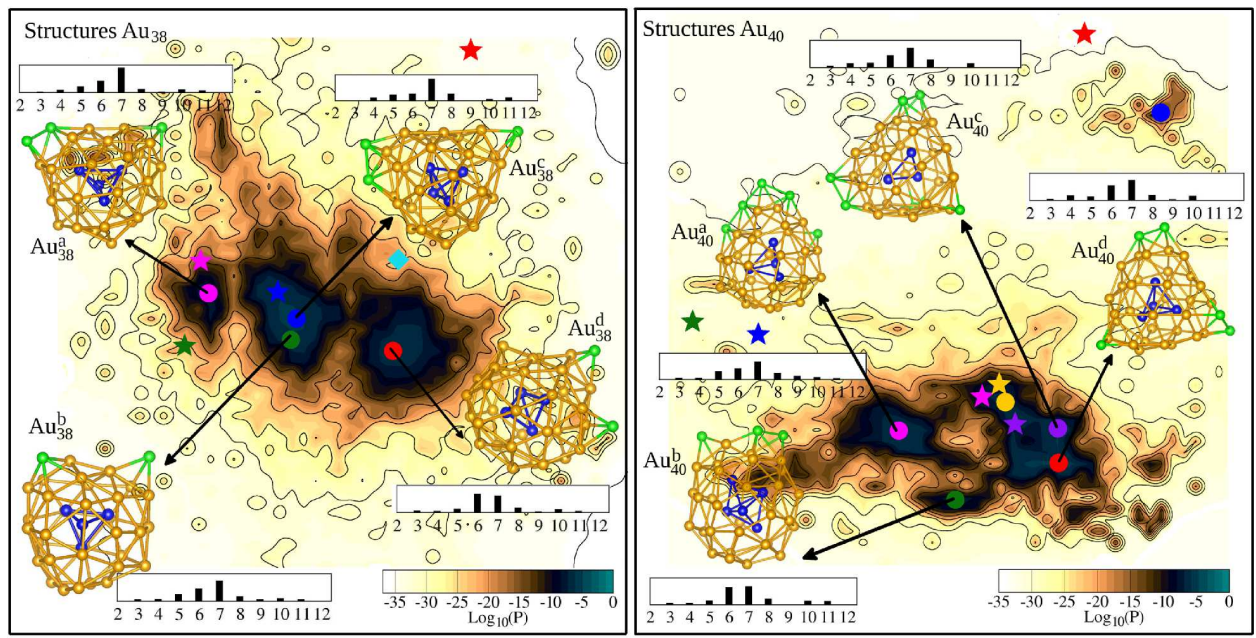}
\caption{Probability of clusters configurations at \SI{300}{\kelvin}, calculated via MBAR as a function of the sketch-map coordinates for the population sampled for \ce{Au38} (left side, based on a total of \SI{400}{\pico\second} of MD sampling) and \ce{Au40} (right side, based on \SI{500}{\pico\second} MD). The selected structures distinguish core (blue), shell (yellow), and protrusions (green) regions. For each selected structure, also the corresponding 13-dimensional coordination-histogram --- the input for the sketch-map dimensionality reduction --- is shown. The dots mark structures selected along the MD runs, positioned in the areas of highest probability and the stars locate the corresponding geometrically relaxed structures (dot/star of the same color indicate initial/final structures for a geometry relaxation). For \ce{Au38}, the pink star is the pGMC found by us (see its structure on Fig. S15). The pGMC identified in Ref. \citenum{Jiang_193402_2011} is marked by a diamond. For \ce{Au40}, the red star is the pGMC found by us, which is also the pGMC identified in Ref. \citenum{Jiang_193402_2011}.  
}
\label{fes_Au38_Au40}
\end{figure*}

The (center-of-mass centered) radial-distribution function $g(r)$ is the histogram in which bin $i$ counts how many atoms are found
between distance $r_{i-1}$ and $r_i$ from the center of mass of the cluster, averaged over all configurations along a trajectory.
\bibnote{Here, $r_0=0$ and the radial increment $r_{i+1}-r_i= \delta r$ is constant and set to 0.1 \AA. For extended systems, the histogram is typically normalized by the volume of the spherical shell between $r_{i+1}$ and $r_i$, but for finite systems this normalization is not relevant and omitted here.}
This descriptor, invariant with respect to rigid rotations and translations of the whole cluster and to permutations of atom identities, gives direct and statistical access to 
the radially layered structure of the clusters, and allows for a parameter-free definition of the core and shell regions.
Fig. \ref{hirshfeld_radial_number}a shows representative $g(r)$s for one trajectory of \ce{Au38} and one trajectory of \ce{Au40}. At both cluster sizes, three clear regions can be distinguished. Some gold atoms are found in the range 1.0--2.5 \AA. These are the atoms constituting the {\em core}. Another group of atoms is found in the range 2.5--5.5 \AA. These are the atoms constituting the {\em shell}. A third group of atoms appears at larger distance. A minimum in $g(r)$ at $r \sim 5.1$ \AA~(\ce{Au38}) and $r \sim 5.5$ \AA~(\ce{Au40)} suggests that atoms contributing to the outer part of the distribution form a well-separated group from the shell atoms. Visual inspection of the cluster structures (e.g., green-colored spheres in Fig. \ref{fes_Au38_Au40}) reveals that these are lower coordinated atoms ``protruding'' from an otherwise triangular-lattice shell, the latter typically comprised of 6--7 fold coordinated atoms. We therefore name this outer region {\em protrusions}. The $g(r)$ for all trajectories, classified also by temperature, are shown in Figs S2--S4 in the \SM. For \ce{Au38} and \ce{Au40}, the three-regions feature is observed for all trajectories and temperatures, with cutoff radii at 2.5 \AA~ for core-shell and in the range 5.0--5.5 \AA~ for shell-protrusions.
For \ce{Au25}, only a (one-atom) core region and a shell region (the latter in the range 2.5--5.5 \AA, as for the larger clusters) is observed.

We note here that atoms are observed to change region along the MD trajectories, so their assignment to a region is dynamical and solely based on the distance from the center of mass. This flexible definition is crucial for our subsequent analysis. In Fig. \ref{hirshfeld_radial_number}, we also show the time evolution of the Hirshfeld charges\cite{Hirshfeld_129_1977} of the core and shell+protrusion regions (panels (b) and (e)), and of the number of atoms belonging to these regions (panels (c) and (f)), for \ce{Au38} and \ce{Au40}. These observables are reported for all trajectories in Figs S2--S5 of the \SM. The qualitative behavior is remarkably consistent at all cluster sizes and temperatures: the core is always fractionally cationic, while the shell+protrusions is always fractionally anionic. 

The number of atoms in the core is very stable. For \ce{Au25} it is always one and for \ce{Au38} it is almost always four, with infrequent fluctuation to five (in the trajectories that started from the structure with a 5-atom core, one atom quickly -- in a few picoseconds -- moves to the shell region). For \ce{Au40}, the behavior is more complex. Trajectories starting with four atoms in the core maintain a 4-atom core (with sporadic fluctuations to five) throughout the time evolution, even at the higher temperature (\SI{600}{\kelvin}). However, the trajectory starting from a three-atom core (structure `J2' in Fig. \ref{initial_structures}) fluctuates between three and four atoms in the core for the \SI{50}{\pico\second} at \SI{600}{\kelvin} and remains with a 3-atom core during the subsequent \SI{50}{\pico\second} time interval at \SI{300}{\kelvin}, indicating a kinetic trapping in a metastable state. When starting with a 5-atom core (EAM-RBHMC pGMC structure), it takes \SI{500}{\kelvin} and more than \SI{10}{\pico\second} to see the \TBP\ 5-atom core transform into a tetrahedral 4-atom core. At \SI{600}{\kelvin}, the transformation is faster (fraction of a picosecond, but followed by frequent fluctuation back to a 5-atom core) while at \SI{400}{\kelvin} the 5-atom core remains stable. The subsequent parts of the trajectories at \SI{300}{\kelvin} maintain the number of atoms in the core, as at the higher temperature. In this case, we have also the signature of a kinetic trapping that is easily overcome at \SI{600}{\kelvin}.

The number of atoms belonging to either shell or protrusions regions (for \ce{Au38} and \ce{Au40}) is quite stable, but with larger fluctuations ($\pm$ 2--3 atoms around the average value along the trajectories). Average sizes of the protrusions regions range from to 2 to 7 atoms, depending on the cluster size and the temperature. 

\begin{figure}[t!]
\centering
\includegraphics[width=0.45\textwidth]{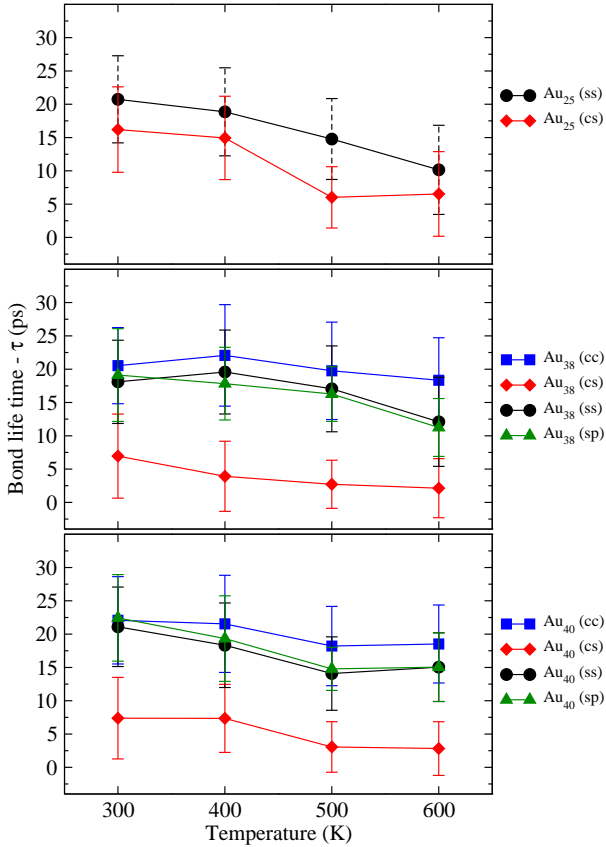}
\caption{(a) Average bond lifetime, $\tau$, and sample standard deviation (bars) 
among the core-core atoms ($\tau_{cc}$), core-shell atoms ($\tau_{cs}$), shell-shell 
atoms ($\tau_{ss}$), and shell-protrusions ($\tau_{ps}$) atoms, for the \ce{Au25}, 
\ce{Au38}, and \ce{Au40} nanoclusters as function of the sampled temperatures. The lines are a guide to the eye.
}
\label{bond_life_time}
\end{figure}

In order to further analyze the nature of the different regions in the clusters, we measured the average bond lifetimes, according to whether the bonded pair belongs to one region or is across regions. To this purpose, we first define a bonded pair (at a given time step) as any pair of atoms at distance smaller than \SI{3.6}{\AA}. This cutoff value is assigned as the first minimum of the the pair-distribution function.\bibnote{It is a concept similar to the $g(r)$ defined above, but here the distance between any pair of atoms is binned. The position of the first minimum (see Fig. S6 in the \SM) is relatively independent of cluster size, temperature, and region to which the atoms belong.} Next, if a bond is found at any time-step, it is monitored until it is first broken (i.e., until the distance between the two atoms exceeds the threshold): The elapsed time is the bond lifetime and it is binned considering whether both atoms are in the core, or one in the core and one in the shell, etc. In Fig. \ref{bond_life_time} we show $\tau$, the region-dependent average value of the bond lifetime, at each cluster size and temperature (if more trajectories sampled the same temperature, we averaged over them). We also show the sample standard deviation of each bond-lifetime histogram as an error-bar, in order to quantify the spread of this quantity under the different conditions. More details on the bond lifetimes, broken down to single trajectories, are given in Table S2 of the \SM. We observe a striking difference in behavior between bonds shared by atoms both belonging to the core, or both belonging to the shell, and bonds across core and shell. The latter are systematically at least 3--4 times shorter lived than the former. The shell-protrusions bonds behave similarly to shell-shell bonds, suggesting that the two regions are structurally separated but dynamically very similar.\bibnote{Protrusions-protrusions bonds were too few to yield a robust statistics at all sizes and temperatures} The dynamics of the core-shell bonds suggests a picture of a shell that is loosely bound to the core. Visual inspection of the trajectories of \ce{Au40} and \ce{Au38} confirms that the core atoms behave like a inner cluster ``confined'', and almost freely rotating, inside the shell. In the case of \ce{Au25}, the single-atom core ``rattles'' around the center of mass. Consistently, $g(r)$ for \ce{Au25} (Fig. S3) shows that the one-atom core is positioned on average at about 0.25~\AA~from the center of mass.

In order to describe the statistical distribution of structures at each cluster size, as sampled along the MD trajectories, we adopted an unsupervised machine-learning approach, namely the nonlinear dimensionality-reduction algorithm termed {\em sketch-map}, as introduced in Ref. \citenum{Ceriotti_13023_2011}. First, similarly to Ref. \citenum{Ceriotti_1521_2013}, we describe each configuration along a MD trajectory with a coordination histogram, where the $j$-th bin is the number of atoms that have coordination $j$, i.e., that have $j$ atoms at a distance within the same threshold defined above for the bond breaking. A graphical example on how a cluster structure is mapped into a coordination histogram is given in Fig. S7 of the \SM. The histogram is normalized, i.e., each bin is divided by the number of atoms in the cluster. For the clusters considered, we found that the maximum coordination is 12, and each histogram was therefore set to have dimension 13, starting from 0-fold coordinated atoms (i.e., atoms evaporated from the clusters) which were, however, not observed in the present study.
This 13-dimensional representation is already a dimensionality reduction from the 3$N$ coordinates for a $N$-atom cluster. Such representation is roto-translational and permutation invariant, similarly to $g(r)$. In this preliminary dimensionality reduction, some information is necessarily lost, as two different clusters may accidentally have the same coordination histogram. Nonetheless, the coordination histogram captures essential information regarding the bonding network for the clusters. Next, we define a distance in the coordination-histogram representation, which is simply the Euclidean distance between histograms, considered as vectors. The sketch-map algorithm \cite{Ceriotti_13023_2011} then embeds, in 2-dimensions, the high dimensional points (here, in the coordination-histogram representation), by representing points that are close (distant) in high dimension as close (distant) in the 2-dimensional map. 
The axes in the plots do not have any physical meaning and cannot even be labelled explicitly, what matters are the distances. Still, each high-dimensional point is mapped into a specific point on the 2-d map. Therefore, by discretizing the 2-d coordinates, one can count how many points land in each discrete (2-d) bin and color-code the map according to the number of points in each bin. For better statistics, we adopted a Boltzmann re-weighting approach, the multistate Bennet acceptance-ratio (MBAR) algorithm,\cite{MBAR}
to estimate the probability that a structure falls in a given bin, by using the trajectories at all temperatures, while the shown plot refers to \SI{300}{\kelvin} only. In this way, the above-described kinetic-trapping issues at \SI{300}{\kelvin} are mitigated by simultaneous use of trajectories at higher temperatures, where no kinetic trapping is observed.
For a converged sampling, the logarithm of these probabilities is proportional to the negative of the free-energy of the given Au$_N$ cluster. In our case, however, the samplings are most likely not converged 
and the probabilities are therefore only qualitative.
The results for \ce{Au38} and \ce{Au40} are shown in Fig. \ref{fes_Au38_Au40} (results for \ce{Au25} are in Fig. S8 of the \SM). 
Absolute errors for the same sketch-maps, also evaluated via MBAR, are in Figs. S9--S11 of the \SM. 

Since the positioning of the data points on the sketch-maps is determined only by their 13-dimensional coordination-histogram representation, the criterion of the color coding can be varied at fixed sketch-map, to highlight different aspects of the phase-space sampling. In Figs. S12--S14 of the \SM, we assign one color in the sketch-maps for data points belonging the same MD trajectory, to show that the trajectories all diffuse over broad regions of the sketch-map and therefore of the configurational space.

For \ce{Au38} (Fig. \ref{fes_Au38_Au40}, left), we observe three rather distinct configurational basins. The structures belonging to the left and center basins have a majority of 7-fold coordinated atoms, while for the right basin, structures have a similarly high population of 6- and 7-fold coordinated atoms. The difference between the left and center basins is that structures in the center basin are more spherically shaped, while structures in the left basin are more distorted, with locally planar portions of the shell. Incidentally, both the RBHMC and `J' initial structures belong to the center basin. Furthermore, structures similar to the `J' structure are visited during the MD trajectories started from the RBHMC structure, suggesting a rather broad sampling of the configurational space along the MD trajectories. The connections between the three basins are narrow funnels, which means that, during the MD trajectories, while structures diffuse freely inside the basins, they need specific, coordinated, and infrequent structural rearrangements in order to pass from one basin to another.
The power of sketch-map is to reveal this latent order, invisible to human eye in the high-dimensional representation, even though there might not be a simple description in terms of geometrical features of the structures.

For \ce{Au40} (Fig. \ref{fes_Au38_Au40}, right), we also observe three basins, albeit less well-defined than for \ce{Au38}. The center-bottom basin contains structures with a \TBP\ 5-atom core and a preference for 6- and 7-fold coordinated atoms. The left basin has structures with a tetrahedral 4-atom core, while in the right basins there is a combination of structures with 3- and 4-atom cores, where the common structural motif (of the whole cluster) is the twisted trigonal pyramid. Structures in both the left- and right basins have a majority of 7-fold coordinated atoms. 

For \ce{Au25} (see Fig. S6 in the \SM), we found tubular-like structures (always with one atom in the core) with one side distorted, an aspect reminiscent of the known Au$_{24}^{-}$ pGMC.\cite{Wang_4038_2012} In a different basin, \ce{Au25} shows a further distortion of the tubular structure, yielding locally planar motifs.

The sketch-maps in Fig.~\ref{fes_Au38_Au40} (and Fig. S6) report also the position of the pGMC and other local minima. We note that these structures are not located in the main basins, which means that these structures are not in the regions of the configurational space mostly visited by the MD sampling. Keeping in mind the caveats expressed above, we refer to these ``darker'', most visited regions, as low-free-energy regions in the following.
To rationalize the finding that low-energy, relaxed structures deviate significantly from the low-free-energy configurations, we selected from the MD trajectories several structures in the lowest-free-energy areas of the sketch-maps and geometrically relaxed them to find their local minimum energy structures. This is shown in Fig. \ref{fes_Au38_Au40} (and Fig. S6) with differently colored dots (selected structures from the MD sampling) and stars (relaxed structures), where dot-star pairs with the same color depict initial and final structures for one relaxation. 
In all cases, the relaxed structures are in higher-free-energy regions compared to the selected initial structures. Interestingly, one of the selected structures from the high-probability regions, labeled as 40d, relaxes into the pGMC, further supporting the observation that high-probability structures at finite temperature are significantly different, based on their coordination histograms, from the pGMC as well as from the other lower-energy local minima. 
We inspected the change upon relaxation of $(i)$ gyration radius, $(ii)$ number of core-shell bonds, and $(iii)$ the average effective coordination number\cite{Hoppe_25_1970,DaSilva_023502_2011} (ECN).\bibnote{The ECN is an indicator of the average coordination of atoms in a cluster. It adopts a self-consistent, parameter-free, smoothly decaying definition of bond and therefore coordination number for each atom.} These values are reported in Table S3. As general trend, the gyration radius decreases and the number of core-shell bonds increases upon relaxation. At the same time, the ECN {\em always} increases. This indicates that there is a contraction of the clusters upon relaxation and, in particular, a tightening of the core-shell binding. Together with the bond lifetime analysis (Fig. \ref{bond_life_time}), this observation further strengthens the picture of typical gold cluster structures at \SI{300}{\kelvin} and higher temperatures with weak coupling between core and shell and therefore larger configurational freedom (hence lower free energy) than relaxed structures. 

We note, incidentally, that the relaxation of the low-free-energy structures, let us identify pGMC for \ce{Au25} and \ce{Au38}, previously unreported in the literature (pink stars in Fig. \ref{fes_Au38_Au40}, left  and S6). The corresponding structures are shown in Fig. S15. For \ce{Au40}, the pGMC we found (red star in Fig. \ref{fes_Au38_Au40}, right) corresponds with the pGMC in Ref. \citenum{Jiang_193402_2011}.

\begin{figure}[t!]
\centering
\includegraphics[width=0.48\textwidth]{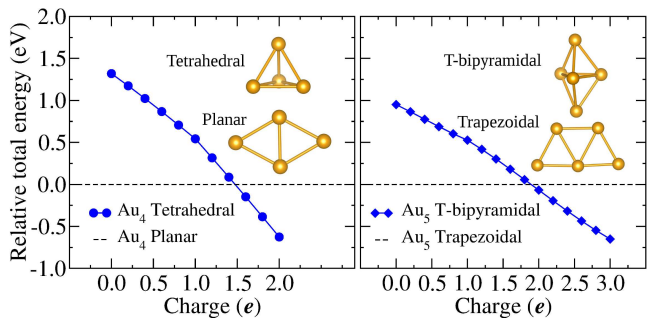}
\caption{Left (Right): Relative stability of the tetrahedral (\TBP) structure with respect to the planar \ce{Au4} (\ce{Au5}) structure, as function of the positive total charge of the cluster.}
\label{relative-energy}
\end{figure}

Finally, we investigated the reason for the stability of three-dimensional 4- and 5-atom cores, while isolated neutral \ce{Au4} and \ce{Au5} are known to be planar. \cite{Ghiringhelli_083003_2013} We measured the amount of positive charge needed to stabilize the 3-d structure vs the planar one for both \ce{Au4} and \ce{Au5}. In Fig. \ref{relative-energy}, we show the energy of the 3-d structure, relative to the planar one for both \ce{Au4} and \ce{Au5}, where both structures are geometrically unrelaxed for each charge.\bibnote{Some structures like the neutral tetrahedron or \TBP\ are mechanically unstable, i.e., have imaginary vibrational frequencies (see Fig. S16 and S17 in the \SM). In Fig. S16 we show the relaxation of the structures by fixing their symmetry and in Fig. S17 without fixing symmetry.} We find that tetrahedral \ce{Au4} becomes more stable than planar \ce{Au4} with a charge of 1.5$e$, while \TBP\ \ce{Au5} needs 2.0$e$ to become more stable than planar \ce{Au5}. These charge values, even without symmetry constraints, are much larger than the small positive charges observed for the core atoms. Therefore, charge alone is not sufficient to stabilize the 3-d structures, leaving confinement and bonding interactions with the shell as likely reasons for the stabilization.

In summary, we have presented a theoretical study of the stability and dynamics of \ce{Au25}, \ce{Au38}, and \ce{Au40} nanoclusters whose configurational space is sampled via Born--Oppenheimer molecular dynamics at various temperatures. We showed that at room and higher temperature (\num{300}, \num{400}, \num{500}, and \SI{600}{\kelvin}) the gold clusters at the chosen sizes exhibit a dynamical core-shell structure, with a cationic core (single \ce{Au} for \ce{Au25} and mainly tetrahedral \ce{Au4} for \ce{Au38} and \ce{Au40}) loosely bound to an outer anionic flexible shell. We also identify, as a persistent feature, a sub-structure in the cluster shells- one triangular-lattice region with 6- and 7-fold coordinated atoms and an outer region with fewer lower-coordinated atoms that we term {\em protrusions}. By using sketch-maps --- two-dimensional embeddings of the coordination-histogram representation of the clusters --- and Boltzmann reweighting, we built approximate free-energy landscapes, at room temperature, of the studied nanoclusters. These show that the global and local minimum energy structures differ significantly from the structures in low free-energy areas, i.e., the most probable at finite temperature.  Obtained by using a suite of statistical tools that can be applied readily to similar systems, our results are able to highlight some typical characteristics of the finite-temperature populations of \ce{Au25}, \ce{Au38}, and \ce{Au40}, providing insights into the stability and dynamics of their core-shell regions.

\providecommand{\latin}[1]{#1}
\makeatletter
\providecommand{\doi}
  {\begingroup\let\do\@makeother\dospecials
  \catcode`\{=1 \catcode`\}=2\doi@aux}
\providecommand{\doi@aux}[1]{\endgroup\texttt{#1}}
\makeatother
\providecommand*\mcitethebibliography{\thebibliography}
\csname @ifundefined\endcsname{endmcitethebibliography}
  {\let\endmcitethebibliography\endthebibliography}{}

\begin{acknowledgement}
We thank Matthias Scheffler for his support and Bryan Goldsmith for a critical reading of the manuscript and insightful suggestions.	
DGS and JDS acknowledge support from FAPESP (S\~ao Paulo Research Foundation, Grant Numbers 2017/11631-2 and 2014/22044-2), Shell and the strategic importance of the support given by ANP (Brazil’s National Oil, Natural Gas and Biofuels Agency) through the R\&D levy regulation.
DGS and JDS also thank the National Council for Scientific and Technological Development (CNPq) and the Coordination for Improvement of Higher Level Education (CAPES) for financial support.
IPH thanks the National Sciences and Engineering Research Council (NSERC) for financial support.
LMG acknowledges for financial support the European Union’s Horizon 2020 research and innovation program {(\#676580: The NOMAD Laboratory --- an European Center of Excellence and \#740233: TEC1p)}, and the Berlin Big-Data Center (BBDC, \#01IS14013E).
Computer time was provided by the Max Planck Computing and Data Facility, the Laboratory of Advanced Scientific Computing (University of S\~ao Paulo), the Department of Information Technology -- Campus S\~ao Carlos, and Compute Canada.
\end{acknowledgement}

\begin{suppinfo}
All the MD-trajectory and geometry-relaxation files are uploaded in the NOMAD Repository ( https://repository.nomad-coe.eu/NomadRepository-1.1 ) and can be inspected and downloaded directly via the DOI: http://dx.doi.org/10.17172/NOMAD/2018.10.30-1.
Extra results are available in the supporting information.
\end{suppinfo}


\begin{mcitethebibliography}{49}
\bibitem[Not()]{Note-1}
Here, $r_0=0$ and the radial increment $r_{i+1}-r_i= \delta r$ is constant and
  set to 0.1 \AA. For extended systems, the histogram is typically normalized
  by the volume of the spherical shell between $r_{i+1}$ and $r_i$, but for
  finite systems this normalization is not relevant and omitted here.\relax
\mciteBstWouldAddEndPunctfalse
\mciteSetBstMidEndSepPunct{\mcitedefaultmidpunct}
{}{\mcitedefaultseppunct}\relax
\EndOfBibitem
\bibitem[Not()]{Note-2}
It is a concept similar to the $g(r)$ defined above, but here the distance
  between any pair of atoms is binned. The position of the first minimum (see
  Fig. S6 in the \SM) is relatively independent of cluster size, temperature,
  and region to which the atoms belong.\relax
\mciteBstWouldAddEndPunctfalse
\mciteSetBstMidEndSepPunct{\mcitedefaultmidpunct}
{}{\mcitedefaultseppunct}\relax
\EndOfBibitem
\bibitem[Not()]{Note-3}
Protrusions-protrusions bonds were too few to yield a robust statistics at all
  sizes and temperatures\relax
\mciteBstWouldAddEndPuncttrue
\mciteSetBstMidEndSepPunct{\mcitedefaultmidpunct}
{\mcitedefaultendpunct}{\mcitedefaultseppunct}\relax
\EndOfBibitem
\bibitem[Not()]{Note-4}
The ECN is an indicator of the average coordination of atoms in a cluster. It
  adopts a self-consistent, parameter-free, smoothly decaying definition of
  bond and therefore coordination number for each atom.\relax
\mciteBstWouldAddEndPunctfalse
\mciteSetBstMidEndSepPunct{\mcitedefaultmidpunct}
{}{\mcitedefaultseppunct}\relax
\EndOfBibitem
\bibitem[Not()]{Note-5}
Some structures like the neutral tetrahedron or \TBP\ are mechanically
  unstable, i.e., have imaginary vibrational frequencies (see Fig. S16 and S17
  in the \SM). In Fig. S16 we show the relaxation of the structures by fixing
  their symmetry and in Fig. S17 without fixing symmetry.\relax
\mciteBstWouldAddEndPunctfalse
\mciteSetBstMidEndSepPunct{\mcitedefaultmidpunct}
{}{\mcitedefaultseppunct}\relax
\EndOfBibitem
\end{mcitethebibliography}
\end{document}